\documentclass[aps,prb,twocolumn,longbibliography]{revtex4-2}
\usepackage{hyperref}
\hypersetup{
    colorlinks = true,
    linkcolor = [rgb]{0.70, 0.13, 0.13},
    citecolor = [rgb]{0.25, 0.41, 0.88},
    urlcolor = [rgb]{0.25, 0.41, 0.88}
}
\usepackage{amsmath}
\usepackage{amsfonts}
\usepackage{amssymb}
\usepackage{pifont}
\usepackage{mathtools}
\usepackage{bm}
\usepackage{cases}
\usepackage{braket}
\usepackage[version=4]{mhchem}
\usepackage{graphicx}
\allowdisplaybreaks%
\usepackage{makecell}
\usepackage{tikz}

\usepackage{blindtext}

\makeatletter
\newcommand*{\sumcirclearrowleft}{%
  \DOTSB
  \mathop{
    \mathchoice
      {\rlap{\kern.25em\rotatebox[origin=c]{-90}{$\circlearrowleft$}}{\sum}}
      {\vcenter{\rlap{\kern.2em\rotatebox[origin=c]{-90}{$\scriptscriptstyle\circlearrowleft$}}}{\sum}}
      {\sum}{\sum}
  }\slimits@
}

\newcommand*{\sumcirclearrowright}{%
  \DOTSB
  \mathop{
    \mathchoice
      {\rlap{\kern.25em\rotatebox[origin=c]{90}{$\circlearrowright$}}{\sum}}
      {\vcenter{\rlap{\kern.2em\rotatebox[origin=c]{90}{$\scriptscriptstyle\circlearrowright$}}}{\sum}}
      {\sum}{\sum}
  }\slimits@
}
\makeatother

\begin{document}

\title{Inelastic neutron scattering on Ce-pyrochlores: Signatures of electric monopoles}

\author{Pengwei Zhao$^{1,2}$}
\author{Gang V. Chen$^{1,2}$}
\email{chenxray@pku.edu.cn}
\affiliation{$^{1}$International Center for Quantum Materials, School of Physics, Peking University, Beijing 100871, China}
\affiliation{$^{2}$Collaborative Innovation Center of Quantum Matter, 100871, Beijing, China}

\begin{abstract}
We revisit the pyrochlore spin liquid materials Ce$_2$Sn$_2$O$_7$ and Ce$_2$Zr$_2$O$_7$
by examining the existing experiments. We continue to rely on the special properties of 
the dipole-octupole nature of the Ce$^{3+}$ moment. The inelastic neutron scattering (INS)
measurement in the octupolar U(1) spin liquid {\sl selects} the (gapped) spinon continuum, 
and thus has suppressed spectral weights below the energy threshold of two spinon gaps. 
This measurement, however, includes all other emergent excitations at lower energies in 
the dipolar U(1) spin liquid, in particular, the gapless photon and the continuum of {\sl the electric monopoles}.    
Although the electric monopole continuum is weakly gapped (compared to the larger spinon gap),    
the energy scale is actually close to the gauge photons, and the spectrum largely overlaps with the photons.
Due to the background dual $\pi$ flux for the electric monopoles, the density of states is enhanced 
at lower energies, creating peak structures. This can be contrasted with the linearly suppressed 
spectral weight of the gauge photons at low energies. We propose that the electric monopole continuum 
should be mostly responsible for the low-energy spectrum in the INS measurement 
in the dipolar U(1) spin liquid. With these understanding and calculation, 
we discuss the available experimental results and predict further experiments 
for Ce$_2$Sn$_2$O$_7$ and Ce$_2$Zr$_2$O$_7$. 
\end{abstract}

\maketitle


There has been a continuous interest in the pyrochlore quantum spin liquid~\cite{Rau_2019,Savary_2016}. 
It is an exotic quantum state that is understood quite well theoretically and 
at the same time has a strong experimental connection with the pyrochlore 
quantum spin ice materials~\cite{PhysRevB.69.064404,PhysRevLett.98.157204,PhysRevX.1.021002}. 
After the intensive exploration of the Tb-pyrochlores, 
Yb-pyrochlores and Pr-pyrochlore~\cite{Rau_2019,Savary_2016},
 the recent activities were about the Ce-based pyrochlore magnets,
 where the Ce$^{3+}$ local moments were identified as the dipole-octupole (DO) 
doublets and related to distinct symmetry-enriched spin liquids~\cite{PhysRevLett.112.167203,PhysRevB.95.041106}. 
There are now three different Ce-based pyrochlore compounds,  
Ce$_2$Sn$_2$O$_7$, Ce$_2$Zr$_2$O$_7$ 
and Ce$_2$Hf$_2$O$_7$~\cite{PhysRevLett.115.097202,2019PhRvL.122r7201G,2019NatPh..15.1052G,2020NatPh..16..546S,2023PhRvB.108q4411B,2023PhRvB.108e4438S,2023arXiv230508261P,2022PhRvX..12b1015S,2024PhRvX..14a1005Y,2024arXiv240208723B,smithSingleCrystalDiffuse2024}.
Most experiments were performed on the first two compounds so far. 
In this work, we review some of the existing experiments,
mostly the inelastic neutron scattering (INS) results, and aim to provide some insights 
from the theoretical perspective to the understanding of the experiments.

Since our major point is about an important corner of the low-energy excitation 
spectra that distinguishes the dipolar U(1) spin liquid and the octupolar U(1) spin liquid, 
we simply state the results here. 
The U(1) spin liquid on the pyrochlore lattice is effectively described by the compact quantum electrodynamics, 
which hosts three different excitations. Spinons are sources of the emergent magnetic field, 
while electric monopoles are sources of the emergent electric field. 
The dynamics of the electrodynamics field manifests as photons.
The INS measurement
could detect all the emergent excitations for the dipolar U(1) spin liquid. In particular, due to
the intrinsic $\pi$ dual U(1) gauge flux, the electric monopole continuum stands out 
at low energies and creates the peak structures on top of the suppressed photon intensity. 
In reality, the whole INS spectrum contains the peaks 
from the electric monopole continuum at low energies and the peaks 
from the spinon continuum at higher energies, in addition to the linearly 
suppressed photon modes~\cite{PhysRevLett.108.037202,PhysRevB.86.075154} at low energies. 
The presence of the electric monopole continuum 
for the dipolar U(1) spin liquid for the DO doublets
in the INS measurement were actually claimed 
in our previous work~\cite{PhysRevB.96.195127,PhysRevResearch.2.013334}. 
Because many results were scattered in different parts in the previous works, 
the relation with the later experimental results has not yet been fully synthesized and clearly made.
To better serve the community, 
we fulfill this task by reviewing the existing experiments and performing expanded calculations 
based on the existing experimental progress. 
Arriving these results requires the combination of the microscopic nature of the DO doublet,
the effective spin model, and the experimental measurement. To make the presentation 
of this work self-contained, we will start from the basics before embarking 
on a bit new outcomes.

\begin{table*}
\begin{tabular}{p{7cm}p{9.5cm}}
\hline\hline
Different U(1) QSLs & \makecell{Inelastic neutron scattering measurement}  \\
\hline
{Octupolar U(1) QSL for DO doublets} &  \makecell{Gapped spinon continuum} \\
\hline
{Dipolar U(1) QSL for DO doublets} & \makecell{Gapless gauge photon, {\sl gapped electric monopole continuum}$^{\ast}$ \\and gapped spinon continuum}   
\\ 
\hline
Dipolar U(1) QSL for non-Kramers doublets & \makecell{Gapless gauge photon, gapped electric monopole continuum}   \\ 
\hline
Dipolar U(1) QSL for usual Kramers doublets & \makecell{Gapless gauge photon, gapped electric monopole continuum \\and gapped spinon continuum}  
\\
\hline\hline 
\end{tabular}
\caption{List of different U(1) QSLs for their inelastic neutron scattering properties. 
Here ``usual Kramers doublet'' refers to the Kramers doublet that differs from a DO doublet. 
This is an improved version of Tab.I in Ref.~\onlinecite{PhysRevB.95.041106} by incorporating 
the understanding of Refs.~\onlinecite{PhysRevResearch.2.013334} and \onlinecite{PhysRevB.96.195127}. 
The electric monopole was referred as ``magnetic monopole'' in Refs.~\onlinecite{PhysRevB.69.064404,PhysRevB.94.205107,PhysRevB.96.195127}. 
The point at $\ast$ is what we emphasize and distinguish from the interpretation 
in Ref.~\onlinecite{gao2024emergent}. 
}
\label{tab1}
\end{table*}

We begin with the dipole-octupole (DO) doublet on the pyrochlore lattice. 
Each state of this doublet is a one-dimensional irreducible representation 
of the $D_{3d}$ point group, and is transformed into the other 
by the time reversal operation~\cite{PhysRevLett.112.167203}. 
Thus, they are special type of Kramers' doublet 
whose degeneracy is protected by the time reversal symmetry. More specially, 
their wavefunctions are linear superpositions of the local ${J^z = 3m/2}$ 
(with $m$ an odd integer) where the $z$ direction is along the local 111 direction 
at each magnetic ion. As it is demanded by the point group symmetry, 
this doublet generally occurs among the crystal field states of all spin-orbit-coupled 
local-$J$ moments with ${J > 1}$.  If the local moment $J$ happens to be an odd 
multiplier of 3/2 such as ${J=9/2}$ for Nd$^{3+}$ 
in Nd$_2$Zr$_2$O$_7$~\cite{PhysRevLett.124.097203,2019PhRvB..99n4420X} 
and 
Nd$_2$Sn$_2$O$_7$~\cite{PhysRevB.92.144423,PhysRevResearch.5.L032027} 
and ${J=15/2}$ for Dy$^{3+}$ in Dy$_2$Ti$_2$O$_7$~\cite{PhysRevLett.112.167203}, 
an easy-axis anisotropy would favor the DO doublet as the crystal field ground state. 
For others $J$'s, this way to obtain the DO doublet as the crystal field ground state 
does not seem to apply. 

We are particularly interested in the Ce-pyrochlores with ${J=5/2}$.
The possible spin liquid was first experimentally proposed for Ce$_2$Sn$_2$O$_7$~\cite{PhysRevLett.115.097202}. 
The DO doublet nature of ground state doublet for the Ce$^{3+}$ ion was clarified 
in the theoretical work of Ref.~\onlinecite{PhysRevB.95.041106}, 
as well as the connection to the pyrochlore ice U(1) spin liquids. 
The key insight of the {\sl selective} spectroscopic measurements 
for the spinon continuum in the octupolar U(1) spin liquid 
by the INS spectroscopy
was also made over there~\cite{PhysRevB.95.041106}. 
Later developments were achieved for both Ce$_2$Zr$_2$O$_7$ and Ce$_2$Sn$_2$O$_7$, 
especially Ce$_2$Zr$_2$O$_7$ due to the single-crystal 
samples~\cite{PhysRevLett.115.097202,2019PhRvL.122r7201G,2019NatPh..15.1052G,2020NatPh..16..546S,2023PhRvB.108q4411B,2023PhRvB.108e4438S,2023arXiv230508261P,2022PhRvX..12b1015S,poree2024fractional,2024PhRvX..14a1005Y,2024arXiv240208723B}. 
The relevance of the $\pi$-flux U(1) spin liquid (or U(1)$_\pi$ spin liquid)  
was then raised, and the translational symmetry enrichment on top of 
the point-group symmetry enrichment, and the symmetry fractionalization 
were discussed~\cite{PhysRevResearch.2.013334}. 


Later theoretical works have also made progress to examine the distinction 
between different symmetry enriched 
U(1) spin liquids~\cite{2020PhRvR...2b3253P,2022PhRvB.105c5149D,2023PhRvB.107f4404D,2024PhRvL.132f6502D,2024arXiv240109551D}. The recent thermodynamics and model calculation were even able to extract the parameters 
of the exchange couplings~\cite{Bhardwaj_2022,2024PhRvX..14a1005Y}. 
The results are different for Ce$_2$Zr$_2$O$_7$ and Ce$_2$Sn$_2$O$_7$ that are 
likely located in different U(1) spin liquid ground states of the underlying XYZ spin model. 
More recent INS measurements indeed found different
excitation spectra for Ce$_2$Zr$_2$O$_7$ and Ce$_2$Sn$_2$O$_7$~\cite{poree2024fractional,gao2024emergent}. 
Since very low-temperature thermodynamics may contain some uncertainty in the measurements and fitting,
we rely mostly on the neutron scattering measurements for the major reasoning. 
To understand the INS experiments, we start from the spin model for the Ce-pyrochlores~\cite{PhysRevLett.112.167203,PhysRevB.95.041106}, 
\begin{eqnarray}
H &=& \sum_{\langle ij \rangle} \big[ J_x \tau^x_i \tau^x_j + J_z \tau^z_i \tau^z_j + J_{xz} (\tau^x_i \tau^z_j + \tau^z_i \tau^x_j ) 
\nonumber \\
&& \quad\quad  + J_y \tau^y_i \tau^y_j \big] -\sum_i  (\hat{n} \cdot \hat{z}_i) h\,  \tau_i^z ,
\label{model1}
\end{eqnarray}
where $\tau^x$ and $\tau^z$ transform as a magnetic dipole moment, $\tau^y$ transforms a magnetic octupole moment,
 $h\hat{n}$ is the external magnetic field that only couples to $\tau^z$ linearly, and $\hat{z}_i$ is the local $z$ direction. 
 The crossing $J_{xz}$ term can be eliminated by a rotation around $y$ direction in the pseudospin-$\tau$ 
 space with $S^x_i = \tau^x_i \cos \theta - \tau^z_i \sin \theta, S^z_i = \tau^x_i \sin \theta  + \tau^z_i \cos \theta, 
 S^y_i= \tau^y_i$, and the resulting spin model is of the XYZ form~\cite{PhysRevLett.112.167203,PhysRevB.95.041106}, 
 \begin{eqnarray}
 H_{\text{XYZ}}  &=& \sum_{\langle ij \rangle} \tilde{J}_x S^x_i S^x_j + \tilde{J}_y S^y_i S^y_j + \tilde{J}_z S^z_i S^z_j 
 \nonumber \\     
 && -\sum_i  (\hat{n} \cdot \hat{z}_i) h\,  ( \cos \theta S^z_i -  \sin \theta S^x_i). 
 \label{modelxyz}
 \end{eqnarray}
where $\tilde{J}$'s are related to the $J$'s via the pseudospin rotation.

 \noindent{\emph{INS spectroscopy.}}---Based on the qualitative and powerful symmetry reasoning, 
 we know that, the model in Eq.~\eqref{modelxyz} 
without the field has two symmetry-enriched U(1) spin liquids from the perspective of the point group. 
For a dominant and antiferromagnetic $\tilde{J}_z$ or $\tilde{J}_x$, a (conventional) dipolar U(1) spin liquid 
is realized. In contrast, for a dominant and antiferromagnetic $\tilde{J}_y$, a novel octupolar U(1) spin liquid is realized. 
It was observed that~\cite{PhysRevB.95.041106}, because the magnetic field 
or the neutron spin only couples to the transverse 
components relative to $S^y$ in the octupolar U(1) spin liquid, the inelastic neutron scattering {\sl only} 
measures the spinon continuum (see Tab.~\ref{tab1}). 
An energy threshold of two-spinon gap~\cite{Chen_2023} 
is needed in order to observe the spinon continuum in the neutron scattering measurement. 
With this insight, the spinon continuum spectroscopy for both 0-flux and $\pi$-flux octupolar 
U(1) spin liquids were studied~\cite{PhysRevB.95.041106,PhysRevResearch.2.013334}.
Moreover, 
due to this selective linear coupling for the dipolar spin components, 
a regular unpolarized INS may already function as a polarized INS,
and a polarized INS can be a unnecessary luxury.

Let us re-examine the dipolar U(1) spin liquid for a dominant and antiferromagnetic $\tilde{J}_x$. 
If $J_{xz}$ is weak (or equivalently, $\theta$ is small), the behaviors in the neutron scattering 
measurement in this limit should not be very different from the octupolar U(1) spin liquid. 
In this regime, the coupling to the neutron spin is primarily via $S^z$ component, and the 
coupling to $S^x$ is suppressed by $\sin \theta$. 
The intensity of the $S^x$-$S^x$ correlation, that includes the electric monopole continuum and the gauge photon, 
is suppressed by $\sin^2 \theta$. 
The INS measurement
cannot differentiate this dipolar U(1) spin liquid in this regime from the octupolar U(1) spin liquid,
regardless of whether further translational symmetry enrichment with 0 or $\pi$ flux 
for the spinons is considered. From the perspective of $S^z$ component,
the hidden $S^x$ and $S^y$ components actually have the same effect. 
This can be understood via a rotation of the spin ${\boldsymbol S}_i$ around the $z$ direction by $\pi/2$
that switches $S^x$ and $S^z$. For the same reason, thermodynamics like specific heat 
and magnetic susceptibility cannot differentiate them.  
Thus, in the above mentioned measurements, the dipolar U(1) spin liquid 
for a dominant and antiferromagnetic $\tilde{J}_x$
is not very different from the octupolar U(1) spin liquid. 
We note that, in Ref.~\onlinecite{Bhardwaj_2022}, a finite linear Zeeman coupling to $\tau^x$ 
is introduced in Eq.~\eqref{model1}, and this would be more-or-less equivalent to a finite $\theta$. 
The presence of this extra coupling could in principle 
make the experimental behaviors of the above two phases different. Our 
understanding here is restricted to the regime with the vanishing $\tau^x$ Zeeman coupling
of Eq.~\eqref{model1}.

\noindent{\emph{About Ce$_2$Sn$_2$O$_7$.}}---The recent INS measurement in the powder sample of 
Ce$_2$Sn$_2$O$_7$ actually observed a large spectral weight at $\sim 0.045$meV, 
and the spectral weights are gradually suppressed below this peak energy. 
From the above explanation, this is either consistent with the 
expectation from the octupolar U(1) spin liquid, or compatible with the dipolar U(1) spin liquid 
for a dominant and antiferromagnetic $\tilde{J}_x$ and a small $\theta$. 
On the positive side, the measured excitation continuum for both cases is 
interpreted as the spinon continuum for the former, and the predominantly spinon continuum for the latter. 
The further structures in the momentum and/or energy domain of the spinon continuum 
could differentiate the distinct translation symmetry enrichments, i.e. the 0 or $\pi$
gauge flux for the spinons~\cite{PhysRevB.95.041106,PhysRevResearch.2.013334,2024PhRvL.132f6502D}. 
It is tempting to mention that, recent experiments of Ref.~\onlinecite{2024PhRvX..14a1005Y} 
on both single-crystal and powder Ce$_2$Sn$_2$O$_7$ 
samples with a different preparation condition concluded with a dominant and antiferromagnetic $\tilde{J}_x$ 
and an intermediate $\theta$ ($\approx 0.19 \pi$). The authors suggested that, Ce$_2$Sn$_2$O$_7$ 
is located in the dipolar spin ice regime, but the ground state is ordered in the all-in all-out state. 
Even if the ground state is not ordered, an intermediate $\theta$ ($\approx 0.19 \pi$) for a dipolar U(1) spin liquid
with a dominant and antiferromagnetic $\tilde{J}_x$ would imply a significant
spectral weight due to the electric monopole continuum and the gauge photon
at energies below the two-spinon gap. This is clearly different from the inelastic neutron scattering 
result in Ref.~\onlinecite{poree2024fractional}. 
The discrepancy between Ref.~\onlinecite{2024PhRvX..14a1005Y} and Ref.~\onlinecite{poree2024fractional} 
might arise from the sample preparation~\cite{2024PhRvX..14a1005Y}
and remains to be resolved. 

\begin{table}
\begin{tabular}{lccccc}
\hline\hline
Material & $\tilde{J}_x$ & $\tilde{J}_y$ & $\tilde{J}_z$ & $\theta$ & Reference \\
Ce$_2$Sn$_2$O$_7$ & - & 0.05 & - & 0  & Ref.~\onlinecite{2020NatPh..16..546S} \\
Ce$_2$Sn$_2$O$_7$ & - & 0.048 & - & 0 & Ref.~\onlinecite{poree2024fractional} \\
Ce$_2$Sn$_2$O$_7$ & 0.045 & $-0.001$ &  $-0.012$ & $0.19\pi$ & Ref.~\onlinecite{2024PhRvX..14a1005Y} \\
Ce$_2$Zr$_2$O$_7$ & 0.063 & 0.064& 0.011 & 0 & Ref.~\onlinecite{2022PhRvX..12b1015S}\\
Ce$_2$Zr$_2$O$_7$ & 0.0385 &  0.088 &  0.020 & 0 & Ref.~\onlinecite{Bhardwaj_2022}  \\
Ce$_2$Zr$_2$O$_7$ & 0.076 & - &-& $0.12\pi$ & Ref.~\onlinecite{gao2024emergent} \\
\hline\hline
\end{tabular}
\caption{The fitted exchange parameters from different works. 
In Ref.~\onlinecite{2020NatPh..16..546S}, the authors only obtained $J^{\pm}/\tilde{J}_y \approx -0.015$ where $J^{\pm}$
is defined with respect to the $y$ component with $J^{\pm}= -(\tilde{J}_z+\tilde{J}_x)/4$. 
Likewise, Ref.~\onlinecite{poree2024fractional} obtained $J^{\pm}/\tilde{J}_y \approx -0.11$.
In Ref.~\onlinecite{2022PhRvX..12b1015S}, an equal fitting is obtained when the values of $\tilde{J}_x$ and $\tilde{J}_y$ are switched. 
In Ref.~\onlinecite{Bhardwaj_2022}, a weak but finite linear coupling to $\tau^x_i$ was introduced in Eq.~\eqref{model1}. 
In Ref.~\onlinecite{gao2024emergent}, the authors only obtained $J^{\pm}/\tilde{J}_x \approx -0.28$ where $J^{\pm}$
is defined with respect to the $x$ component with $J^{\pm}= -(\tilde{J}_y+\tilde{J}_z)/4$. 
The energy unit is given in meV. 
}
\label{tab2}
\end{table}

\noindent{\emph{About Ce$_2$Zr$_2$O$_7$.}}---In contrast, the recent INS measurement in the single-crystal sample   
of Ce$_2$Zr$_2$O$_7$ observed enhanced spectral weights at low energies, and the large 
spectral intensity extends down to zero energy within the experimental reach~\cite{gao2024emergent}. 
Without invoking any other features from the experiments, we think this piece of
experimental information alone can be understood from the following three possibilities. 

The first possibility is that, the low-energy spectral weights arise from the spinon continuum with 
a small gap, and the system is either in the octupolar U(1) spin liquid or in the dipolar U(1) spin liquid 
for a dominant and antiferromagnetic $\tilde{J}_x$ and a small $\theta$. 
If the fitted exchange parameters in Tab.~\ref{tab2} are used, however, the system is clearly
not in the regime with weakly gapped spinon excitations. Given the thermodynamic results were used in the fitting, 
 this possibility may be ruled out. 



The second possibility is that, the system is in a dipolar U(1) spin liquid with an intermediate 
$\theta$ such that $\sin^2 \theta$ and $\cos^2 \theta$ are of the same order. 
This can allow for a dominant and antiferromagnetic $\tilde{J}_z$, 
or a dominant and antiferromagnetic $\tilde{J}_x$, or both. 
The third possibility is that, the system is in a dipolar U(1) spin liquid with a dominant and 
antiferromagnetic $\tilde{J}_z$ and a small $\theta$. For the third possibility, the spinon 
continuum at high energies should be suppressed from $\sin^2 \theta$.

For the second and third possibilities, the low-energy spectral weights below the two-spinon 
gap arise from the electric monopole continuum and the gauge photon. But since the spectral 
weight of the gauge photon is intrinsically suppressed at low energies, thus the low-energy 
spectral weights for these two possibilities are mostly given by the electric monopole continuum. 
This is the fundamental difference of our interpretation from Ref.~\onlinecite{gao2024emergent}.

\begin{figure}[t]
    \centering
    \includegraphics[width=0.5\textwidth]{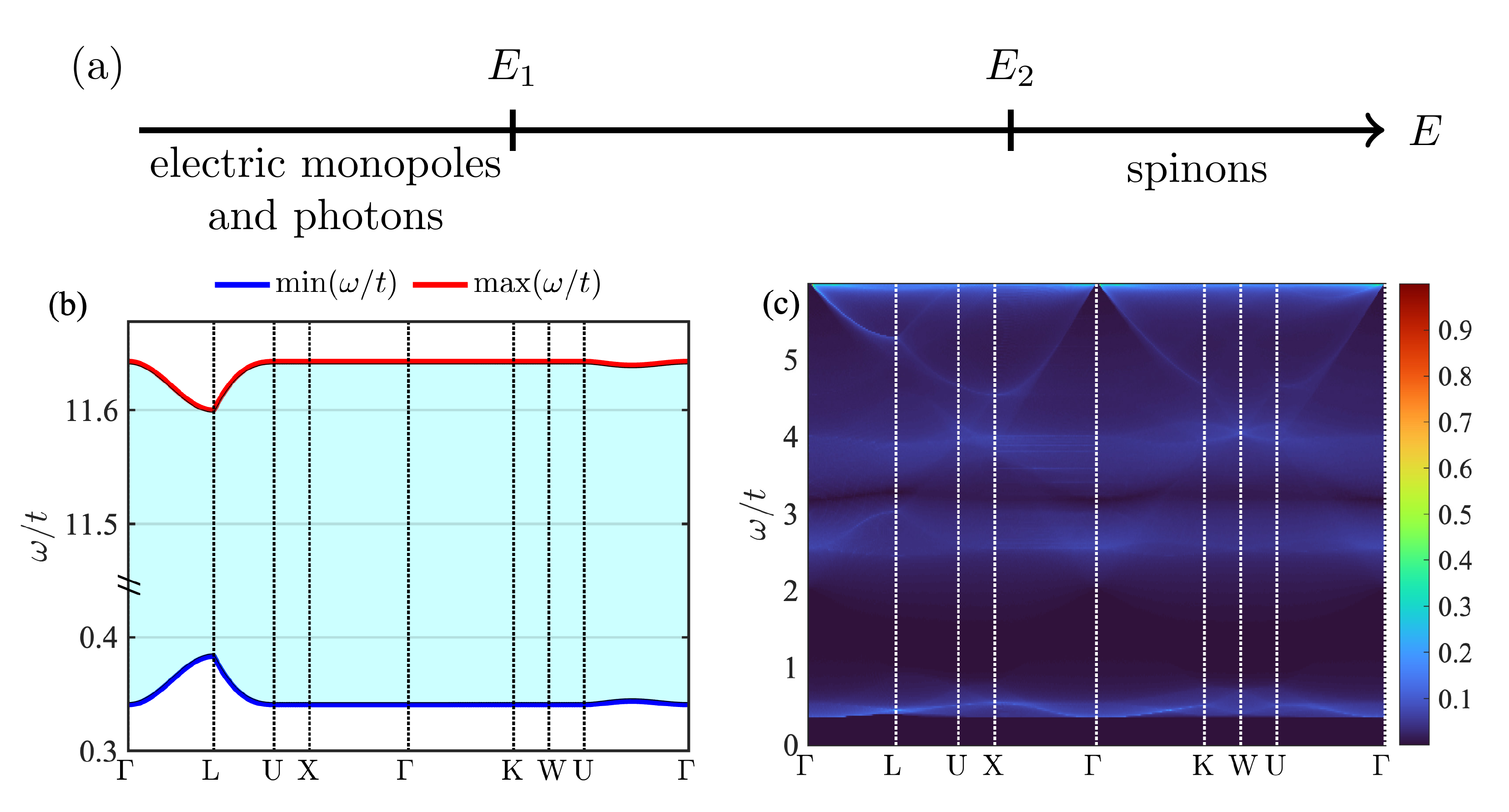}
\caption{(a) The schematic energy level diagram of eletric monopoles, photons and spinons. 
 (b) and (c) are the structure of the two-monopole continuum along the high-symmetry line in the magnetic Brillouin zone: $\Gamma$-$\mathrm{L}$-$\mathrm{U}$-$\mathrm{X}$-$\Gamma$-$\mathrm{K}$-$\mathrm{W}$-$\mathrm{U}$-$\Gamma$. The chemical potential is set to $\mu=-3t$. (b) Lower edge (blue curve) and upper edge (red curve) of the two-monopole continuum. (c) Normalized two-monopole continuum density of states, $\rho^{(2)}(\boldsymbol{k},\omega)$, spanning from the lower edge to half of the continuum width. The brightness indicates the strength of the density of states.}   
 \label{fig:continuum}
\end{figure}

\noindent{\emph{Electric monopole continuum.}}---We here focus on and work out the details 
for the low-energy electric monopole continuum that is shared by these two possibilities. 
Unlike the spinons that hop on the diamond lattice formed by the tetrahedral centers of the 
pyrochlore lattice, the electric monopoles reside on the dual diamond lattice whose bonds
penetrate through the hexagonal plaquette centers of the diamond lattice for the spinons. 
If one sticks to the context of quantum spin ice, the spinons are excitations out of the ice manifold,
while the photon and the electric monopoles are excitations that are built up from the ice manifold
and occur in a much lower energy scale than the spinons (see Fig.~\ref{fig:continuum}). The spinon has a classical analogue 
as the classical magnetic monopole, 
the photon and the electric monopoles are purely of quantum origin and have no classical analogue. 
Moreover, the electric monopoles experience a $\pi$ dual U(1) gauge flux on the dual diamond lattice. 
Thus, the translation symmetry is fractionalized for the electric monopoles.

The Hamiltonian describing the electric monopole hopping on the dual diamond lattice is given as~\cite{PhysRevB.96.195127}, 
\begin{eqnarray}\label{eq:monopole_model}
H_{\text{m}} = -t\sum_{\langle {\boldsymbol r}{\boldsymbol r}' \rangle} e^{-2\pi i\alpha_{\boldsymbol{r} \boldsymbol{r}'}}\Phi_{\boldsymbol{r}}^\dagger\Phi_{\boldsymbol{r}'}^{} - \mu\sum_{\boldsymbol{r}} \Phi_{\boldsymbol{r}}^\dagger\Phi_{\boldsymbol{r}}^{},
\end{eqnarray}
where $\Phi_{\boldsymbol{r}}^\dagger$ ($\Phi_{\boldsymbol{r}}$) represents the monopole creating (annihilation) operator, and
$t$ and $\mu$ denote the hopping and chemical potential of the monopoles, respectively. 
The position $\boldsymbol{r}$ spans the dual diamond lattice, 
comprising the A and B sublattices. The dual gauge link 
$\alpha_{\boldsymbol{r}\boldsymbol{r}'}$, capturing the $\pi$ dual U(1) gauge flux due to the effective spin-1/2 nature of the moment, 
is chosen as
${2\pi \alpha_{\boldsymbol{r},\boldsymbol{r}+\boldsymbol{e}_\mu} = \xi_\mu \boldsymbol{Q}\cdot\boldsymbol{r}}
$, where ${\boldsymbol{r}\in}$ sublattice A and ${\boldsymbol{t}+\boldsymbol{e}_\mu\in}$ sublattice B. 
Here, $\boldsymbol{e}_\mu$ ($\mu=0,1,2,3$) represents the nearest-neighbor vectors 
connecting the two sublattices (see supplementary materials~\cite{SM}). 
To fix the gauge, we choose $(\xi_0,\xi_1,\xi_2,\xi_3)=(0,1,1,0)$ and $\boldsymbol{Q}=2\pi(1,0,0)$. 
Diagonalizing Eq.~\eqref{eq:monopole_model}, we obtain four monopole bands
{\small\begin{eqnarray}
\Omega^{\zeta}_{\eta} = t\zeta\left[4 + 2\eta(3+C_xC_y-C_xC_z+C_yC_z)^{\frac{1}{2}}\right]^{\frac{1}{2}} - \mu,
\end{eqnarray}}
where ${C_\mu=\cos q_\mu}$ ($\mu=x,y,z$), ${\zeta=\pm}$, and ${\eta=\pm}$. Due to the background 
$\pi$ dual U(1) gauge flux, the monopole bands are then defined over a ``magnetic" Brillouin zone that is half
of the crystal Brillouin zone.

\begin{figure}[b]
    \centering
    \includegraphics[width=0.5\textwidth]{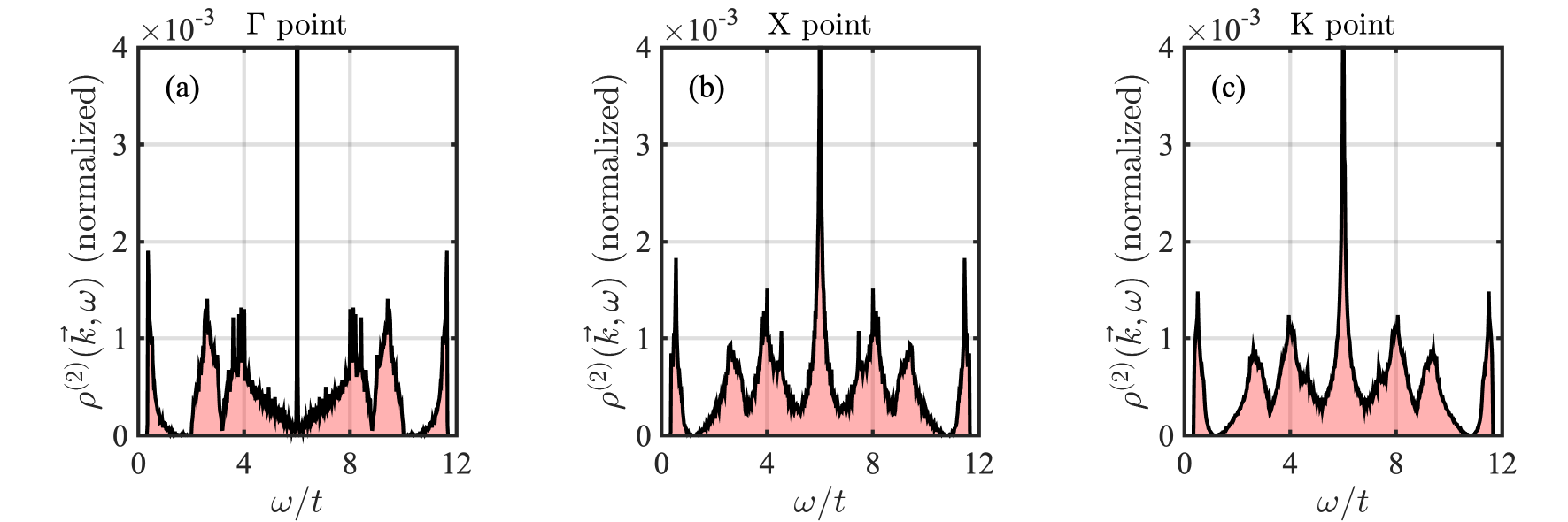}
    \caption{Normalized two-monopole continuum density of states, $\rho^{(2)}(\boldsymbol{k},\omega)$, at three high-symmetry points: $\Gamma$, $\mathrm{X}$, and $\mathrm{K}$. (a) At the $\Gamma$ point, the density of states features a $\delta$-function-like peak at the spectrum center due to the energy-momentum matching of two monopoles. (b) At the $\mathrm{X}$ point and (c) $\mathrm{K}$ point, the density of states exhibits a similar shape, albeit with weaker peaks at the spectrum center compared to the $\Gamma$ point.}
    \label{fig:high_symmetry_points}
\end{figure}

As previously discussed, the two-monopole continuum predominantly contributes to
 the INS signal in the low-energy regime. This signal is expected to correlate 
 with the two-monopole continuum density of states, $\rho^{(2)}(\bm{k},\omega)$, 
 regardless of a specific form factor. 
With the momentum and energy conservation, the density of states is given by
\begin{equation}
    \rho^{(2)}(\boldsymbol{k},\omega) \propto \sum_{\boldsymbol{q},\zeta_1,\zeta_2,\eta_1,\eta_2} \delta\left[\omega - \Omega^{\zeta_1}_{\eta_1}(\boldsymbol{q}) - \Omega^{\zeta_2}_{\eta_2}(\boldsymbol{k} - \boldsymbol{q} - \boldsymbol{Q})\right],
\end{equation}
where $\boldsymbol{k}$ and $\omega$ represent the total momentum and energy of the two monopoles, respectively. 
The offset $\boldsymbol{Q}$ originates from the spatial periodicity of the gauge potential. 
Due to the periodicity of $\rho^{(2)}(\bm{k},\omega)$, however, this offset does not influence the spectrum. 
Specifically, at the $\Gamma$ point, the momenta of the two monopoles cancel out. 
Given the symmetry of the monopole bands under reflection with respect to the ${\Omega=-\mu}$ 
plane and inversion ${\boldsymbol{q}\to-\boldsymbol{q}}$, the continuum spectrum features a flat band near ${\omega=-2\mu}$. 
Consequently, a $\delta$-function-like peak is observable in the two-monopole continuum density of states at the $\Gamma$ point (see Fig.~\ref{fig:high_symmetry_points}(a)).

In the following, we set $t$ as the energy unit and ${\mu=-3t}$ to prevent the electric monopole condensation~\cite{PhysRevB.94.205107}. 
We normalize $\rho^{(2)}(\bm{k},\omega)$ by its maximal value. Given the energy overlapping between the four monopole bands, 
the two-monopole continuum spans continuously from a lower to an upper edge, with a spectrum width of $8\sqrt{2}t$. 
Fig.~\ref{fig:continuum} illustrates the edge shapes along the high-symmetry line 
(for the definition of high-symmetry points, see supplementary materials~\cite{SM}). 
Additionally, Fig.~\ref{fig:high_symmetry_points}(b) and 
(c) display the density of states at two representative high-symmetry points.

%

%












\noindent{\emph{Discussion}}---We discuss the fitted parameters for Ce$_2$Zr$_2$O$_7$ from Tab.~\ref{tab2}. 
For the parameters in Ref.~\onlinecite{2022PhRvX..12b1015S}, since ${\theta \sim 0}$,
as we have discussed, one cannot distinguish the octupolar U(1) spin liquid with a dominant $\tilde{J}_y$
and the dipolar U(1) spin liquid with a dominant $\tilde{J}_x$. 
For both phases, the INS measurement can only probe the spinon continuum at high energies. 
Thus, this set of parameters is inconsistent with the neutron results in Ref.~\onlinecite{gao2024emergent}. 
For the parameters in Ref.~\onlinecite{Bhardwaj_2022}, even though an additional linear Zeeman coupling to $\tau^x$ in Eq.~\eqref{model1}
is introduced, the parameters for the exchange simply favor an octupolar U(1) spin liquid, for which
the INS measurement can only probe the spinon continuum at high energies. 
For the parameters in Ref.~\onlinecite{gao2024emergent}, the system is in the dipolar U(1) spin liquid.
 As $\theta =0.12 \pi$ is in the intermediate regime, all excitations should show up 
in the INS measurement, where the low energy part should be mostly contributed 
by the electric monopole continuum. This is qualitatively compatible with the expectation. 
On a bit more quantitative side, if one uses the fitted couplings of Ref.~\onlinecite{gao2024emergent},
the energy scale for the electric monopoles and photons is approximately set by the ring exchange 
$J_{\text{ring}} = 12 {(J^{\pm})^3/\tilde{J}_x^2}\sim 0.02$meV, while the spinon energy scale is set 
by $\tilde{J}_x$ and renormalized by $J_{\pm}$. Thus, we expect that, 
probably below about 0.05meV, the INS spectrum is contributed by photons and electric monopole continuum, and 
mostly by the electric monopole continuum. The detailed structures of the continuum require a high 
energy resolution to resolve. Otherwise, the continuous excitations are simply crowded up at low energies (see supplementary materials~\cite{SM}).
Roughly above about 0.05meV, the INS spectrum is contributed by spinon continuum combined with other low-energy 
excitations (photon, electric monopoles).

The above analysis, however, has one caveat. If a linear Zeeman coupling to $\tau^x$ 
is introduced in Eq.~\eqref{model1} as Ref.~\onlinecite{Bhardwaj_2022},
 a dipolar U(1) spin liquid with a dominant $\tilde{J}_z$ 
or $\tilde{J}_x$ with $\theta\sim 0$ could not be simply ruled out. About the calculation, the caveat is that
 we treat the electric monopole at the quadratic level and the dual U(1) gauge field 
as a static background. According to Ref.~\onlinecite{PhysRevLett.124.097204}, 
the threshold spinon-pair production could be abruptly enhanced due to the coupling 
to the dynamic gauge field. Similar features would occur for the threshold electric monopole 
production, giving rise to a large density of states at low energies than our current treatment.

It is interesting to vision the effect of a weak magnetic field for a dipolar U(1) spin liquid 
if Ce$_2$Zr$_2$O$_7$ realizes it. 
A weak magnetic field polarizes the effective Ising component and modifies the dual gauge 
flux away from $\pi$ for the electric monopoles. This has two consequences. First, the 
electric monopoles develop Hofstadter-like band structures, and the continuum will have 
more modulations in the energy-momentum domain~\cite{PhysRevB.94.205107}. 
The abundance of the electric monopole continuum at low energy should also be accessible 
through the $\mu$SR measurement at zero/weak field~\cite{2023PhRvB.108q4411B}. 
Second, the electric monopole band will have a nontrivial Berry curvature distribution, generating 
the monopole thermal Hall effect at low temperatures~\cite{PhysRevResearch.2.013066}. 
If Ce$_2$Sn$_2$O$_7$ realizes the octupolar U(1) spin liquid, no such effect is expected. 
In general, the very low energy scale of the monopole physics for the rare-earth magnets makes
the detailed measurement difficult, and this calls for the candidate $d$ electron systems
with higher energy scales and the numerical calculations~\cite{PhysRevResearch.2.042022}. 
So far, one candidate $d$ electron magnets for the DO doublets is Cd$_2$Os$_2$O$_7$, but it is ordered~\cite{PhysRevLett.108.247204}. 


In summary, we propose the low-energy continuous excitations in the recent inelastic neutron scattering 
measurements in Ce$_2$Zr$_2$O$_7$ is mostly contributed from the electric monopole continuum.
Although the gapless gauge phonon is present, its intensity should be suppressed
by at low energies. While the fitted parameters in Ref.~\onlinecite{gao2024emergent} 
are qualitatively compatible with our expectation from the INS scattering 
measurement, our interpretation is different from Ref.~\onlinecite{gao2024emergent}. 
We further demonstrate the structures of the electric monopole continuum in the energy 
and momentum domains.

\emph{Acknowledgments.}---We thank a recent conversation with Bin Gao, Radu Coldea and Bruce Gaulin,
and the previous collaboration with Pengcheng Dai, Brian Maple, Emilia Morosan, Andriy Nevidomskyy and Sang-Wook Cheong.  
This work is supported by the National Science Foundation of China with Grant No.~92065203,
 by the Ministry of Science and Technology of China with Grants No.~2021YFA1400300,
 and by the Fundamental Research Funds for the Central Universities, Peking University.

\bibliography{refs}

\end{document}


\title{Supplementary materials of ``Inelastic neutron scattering on Ce-pyrochlores: Signatures of electric monopoles''}

\author{Pengwei Zhao$^{1,2}$}
\author{Gang V. Chen$^{1,2}$}
\email{gangchen.physics@gmail.com, away from HKU.}
\affiliation{$^{1}$International Center for Quantum Materials, School of Physics, Peking University, Beijing 100871, China}
\affiliation{$^{2}$Collaborative Innovation Center of Quantum Matter, 100871, Beijing, China}

\maketitle

\tableofcontents

\setcounter{table}{0}
\renewcommand{\thetable}{S\arabic{table}}
\setcounter{figure}{0}
\renewcommand{\thefigure}{S\arabic{figure}}
\renewcommand{\theequation}{S\arabic{equation}}

\section{Lattice structure}\label{app:lattice}
The physical spin operators live in a Pyrochlore lattice, which is formed by corner-shared tetrahedra. A diamond lattice can be constructed by connecting the centers of the tetrahedra, on which a U(1) lattice gauge theory is defined. A further duality transformation yields the theory of monopoles, which live in the dual lattice of the diamond lattice: the dual diamond lattice. It consists of two sublattices. The relative vector between these two sublattices is $\boldsymbol{\delta}=\frac{1}{2}(1,1,1)$. Each sublattice is a face-centered cubic (fcc) lattice. Each site of the dual diamond lattice has four nearest neighbors, whose relative vectors are $\boldsymbol{e}_0=\frac{1}{4}(1,1,1)$, $\boldsymbol{e}_1=\frac{1}{4}(1,-1,-1)$, $\boldsymbol{e}_2=\frac{1}{4}(-1,1,-1)$, and $\boldsymbol{e}_3=\frac{1}{4}(-1,-1,1)$. The lattice vectors of the underlying Bravais lattice are chosen to be $\boldsymbol{a}_1=\frac{1}{2}(0,1,1)$, $\boldsymbol{a}_2=\frac{1}{2}(1,0,1)$, and $\boldsymbol{a}_3=\frac{1}{2}(1,1,0)$. Thus, the three lattice vectors of the reciprocal lattice are defined as $\boldsymbol{b}_1=2\pi(-1,1,1)$, $\boldsymbol{b}_2=2\pi(1,-1,1)$, and $\boldsymbol{b}_2=2\pi(1,1,-1)$. As described in the main text, monopole bands in the $\pi$-flux background have an enhanced periodicity, which leads to a smaller Brillouin zone spanned by $\bm{b}_i'=\boldsymbol{b}_i/2$. To reveal the spectrum, it is natural to consider the high-symmetry points defined in Tab.~\ref{tab3} in the smaller Brillouin zone.

\begin{table}[htb]
\begin{tabular}{cc}
\hline\hline
High-symmetry point & Cartesian coordinates \\
\hline
$\Gamma$ & $\pi(0,0,0)$ \\
$\mathrm{X}$ & $\pi(0,0,1)$ \\
$\mathrm{L}$ & $\pi(1/2,1/2,1/2)$ \\
$\mathrm{W}$ & $\pi(0,1/2,1)$ \\
$\mathrm{U}$ & $\pi(1/4,1/4,1)$ \\
$\mathrm{K}$ & $\pi(0,3/4,3/4)$ \\
\hline\hline
\end{tabular}
\caption{The definition of high-symmetry points of the magnetic Brillouin zone.}
\label{tab3}
\end{table}

\section{More details on the spectrum}
\label{sec2}
 
\begin{figure}[h!tb]
    \centering
    \includegraphics[width=0.5\textwidth]{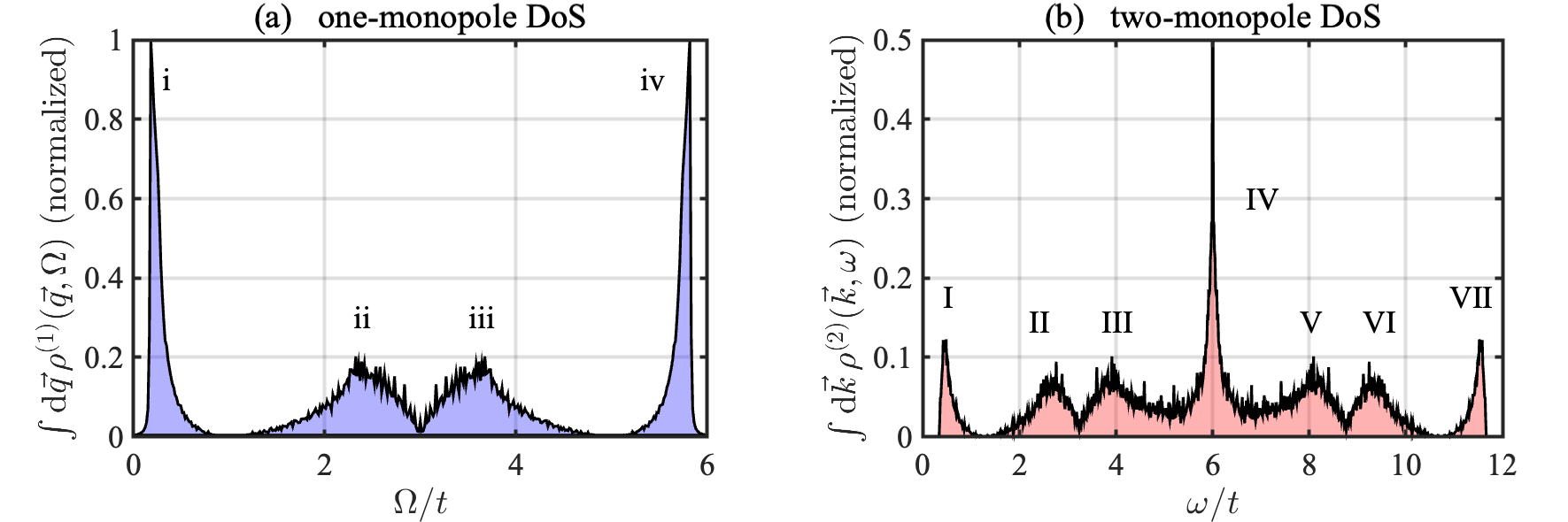}
    \caption{Illustration of the density of states (DoS) for the parameter setting $\mu=-3t$. (a) Normalized momentum-integrated one-monopole DoS, $\int\mathrm{d}\boldsymbol{q},\rho^{(1)}(\boldsymbol{q},\Omega)$, exhibiting four peaks labeled i, ii, iii, and iv, corresponding to the four monopole bands. (b) Normalized momentum-integrated two-monopole DoS, $\int\mathrm{d}\boldsymbol{k},\rho^{(2)}(\boldsymbol{k},\omega)$, showing seven peaks labeled I, II, III, IV, V, VI, and VII. The enhancement of a sharp peak in the middle is prominently visible.}
    \label{fig:dos}
\end{figure}

The peak structure of the two-monopole continuum DoS appears to be insensitive to the total momentum. To validate this observation, we compute the momentum-integrated DoS, given by $\int\mathrm{d}\boldsymbol{k},\rho^{(2)}(\boldsymbol{k},\omega)$. In Figure~\ref{fig:dos}(b), the momentum-integrated DoS exhibits seven peaks, with their positions centered around the highest peak located at $\omega=-2\mu=6$. This peak structure can be elucidated by examining the DoS of the four monopole bands. Each monopole band contributes a peak to the one-monopole DoS, and their collective effect manifests as the observed peak structure in the two-monopole continuum DoS.

\begin{figure}[h!tb]
    \centering
    \includegraphics[width=0.5\textwidth]{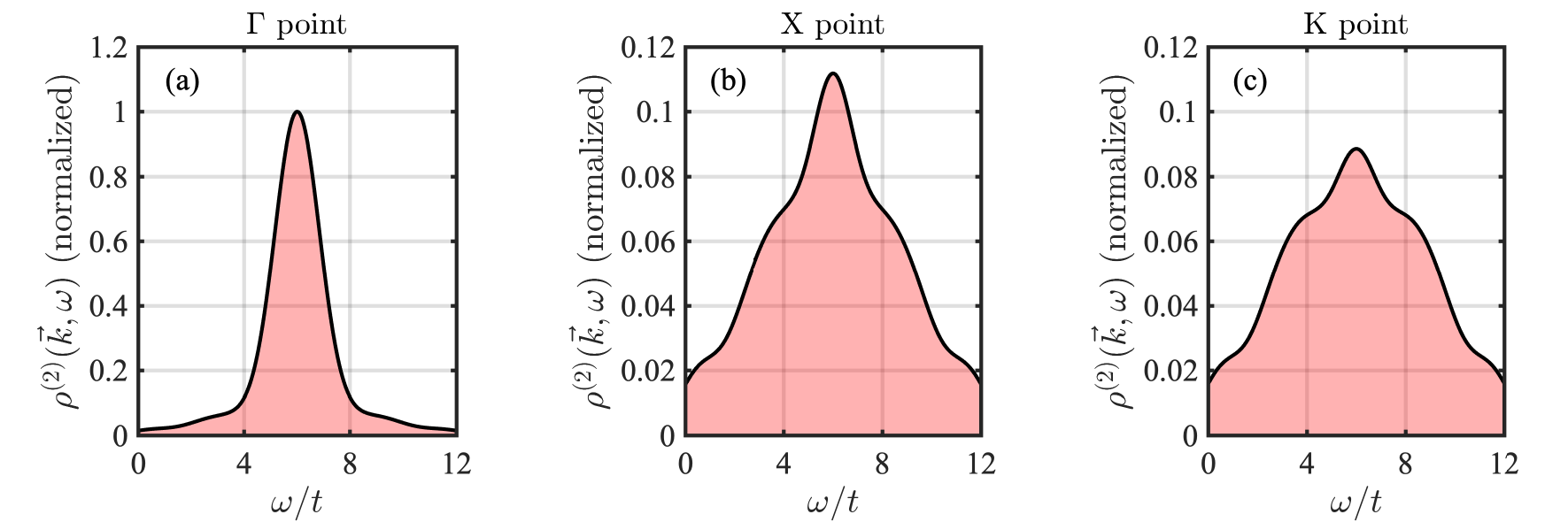}
    \caption{Normalized two-monopole continuum density of states, $\rho^{(2)}(\boldsymbol{k},\omega)$, considering a finite energy resolution at three high-symmetry points: $\Gamma$, $\mathrm{X}$, and $\mathrm{K}$.}
    \label{fig:resolved_dos}
\end{figure}

While the spectrum presents fine structures, these might not be readily discernible in INS experiments. 
Being excitations within the ice manifold, monopoles operate on a relatively small energy scale, 
making their spectrum width potentially compatible with the energy resolution of INS. 
Consequently, a broadened peak is anticipated rather than a sharp one. To account for finite energy resolution, 
we simulate it using a Gaussian function with a standard deviation equal to $10\%$ of the spectrum width. 
Figure~\ref{fig:resolved_dos} displays the continuum density of states at three high-symmetry points: 
$\Gamma$, $\mathrm{X}$, and $\mathrm{K}$.

To detect the electric monopole continuum, one potential strategy could involve synthesizing both an octupolar U(1) 
spin liquid material and a dipolar U(1) spin liquid material with identical parameters (under the corresponding spin component permutation). In the case of the octupolar U(1) spin liquid, 
INS measures only the gapped spinon continuum. By subtracting the spinon continuum signal of the octupolar U(1) spin liquid 
from the INS signal of the dipolar U(1) spin liquid, the remaining signal would predominantly reflect the monopole continuum and gauge photon. 
This differential approach may provide a robust method to isolate and detect the electric monopole continuum signal experimentally.

\section{The derivation of compact U(1) lattice gauge theory}
We provide more details about the construction of a compact U(1) lattice gauge theory (LGT) and the mean-field theory of electric monopoles. We start with the XYZ model
\begin{equation}
    H_{\text{XYZ}}=\sum_{\braket{ij}}\left(\tilde{J}_zS^z_iS^z_j-J^{\pm}S^+_iS^-_j-J^{\pm}S^-_iS^+_j\right).
\end{equation}
$i$ and $j$ label the pyrochlore lattice sites, and $J^{\pm}= -(\tilde{J}_x+\tilde{J}_y)/4$. We consider an easy-axis limit $\tilde{J}_z\gg |J^{\pm}|$. Because pyrochlore lattice is formed by corner-shared tetrahedra, the non-perturbative Hamiltonian is reduced to $H_0=\tilde{J}_z\sum_{\tetrahedron}\left(S^z_{\tetrahedron}\right)^2/2$, where $S^z_{\tetrahedron}=S^z_1+S^z_2+S^z_3+S^z_4$ is the total $S^z$ spin in a tetrahedron $\tetrahedron=\{1,2,3,4\}$. The ground states of $H_0$ are ``two-in-two-out'' states, which form the ice manifold. At this stage, we can understand the first kind of excitations: spinons. Spinons are excitations of $H_0$ out of the ice manifold (see Table~\ref{tab4}). Namely, flipping a $S^z$ spin on a site creates two spinons in the adjacent tetrahedra and increases the energy by $\tilde{J}_z$. Thus, each spinon carries energy quantum $\tilde{J}_z/2$. Other two kinds of excitations, photons, and electric monopoles, appear at the perturbative level of $J^{\pm}$ and they are excitations in the ice manifold, which have lower energy compared to spinons.
The perturbation theory yields the effective Hamiltonian $H_{\text{eff}}=-J_{\text{ring}}\sum_{\hex}\left(S^+_1S^-_2S^+_3S^-_4S^+_5S^-_6+\text{h.c.}\right)$ with $J_{\text{ring}}=12(J^{\pm})^3/(\tilde{J}_z)^2$ and $\hex=\{1,2,3,4,5,6\}$ the hexagon-shaped loop in the pyrochlore lattice. 

\begin{table}[htb]
\begin{tabular}{cccc}
\hline\hline
Notation 1 & Notation 2 & Notation 3 (this paper) & Physical origin\\
\hline
Spinon & Magnetic monopole & Spinon & Gapped excitations out of the ice manifold \\
Magnetic monopole & Electric monopole & Electric monopole & Gapped excitations within the ice manifold \\
Gauge photon & Gauge photon & Gauge photon & Gapless excitations of the lattice electromagnetic field \\

\hline\hline
\end{tabular}
\caption{Three equivalent different notations and the physical origin of excitations in the pyrochlore U(1) QSL. Notation 1 was introduced in Ref.~\cite{PhysRevB.69.064404}. Notation 2 was used by Ref.~\cite{wangTimeReversalSymmetricQuantum2016a}. Our paper adopts notation 3. In our notation, spinons are the gapped excitations out of the ice manifold, while electric monopoles are the gapped excitations within the ice manifold. In the compact U(1) lattice gauge theory of pyrochlore U(1) QSL, spinons are the source of the magnetic field while the electric monopoles are the sources of the electric field. The excitation of the electromagnetic field itself yields the gapless gauge photon.}
\label{tab4}
\end{table}

The effective Hamiltonian is generally interpreted as a compact U(1) lattice gauge theory. 
To present the exact construction, we denote the center of these tetrahedra as $\bm{r}$. 
A diamond lattice can be obtained by connecting the centers of the corner-shared tetrahedra. 
Each pyrochlore lattice site corresponds to a diamond lattice link. 
The diamond lattice consists of sublattice A and sublattice B. 
We define $\bm{e}_\mu$ ($\mu=0,1,2,3$) as the four vectors that connect the sites of sublattice A of the diamond lattice to their nearest neighbors. 
Thus, the four vectors that connect the sites of sublattice B to their nearest neighbors are $-\bm{e}_\mu$. 
Based on the four vectors, we can label the diamond lattice links by $(\bm{r},\bm{r}+\eta_{\bm{r}}\bm{e}_\mu)$, where $\eta_{\bm{r}}=+1$ ($\eta_{\bm{r}}=-1$) if $\bm{r}$ belongs to sublattice A (sublattice B). 
Then, we define an integer-valued magnetic field $B_{\bm{r},\bm{r}+\eta_{\bm{r}}\bm{e}_\mu}$ such that $S^z_i=\eta_{\bm{r}}B_{\bm{r},\bm{r}+\eta_{\bm{r}}\bm{e}_\mu}-1/2$, where $i$ is the pyrochlore lattice site located at the center of $\bm{r}$ and $\bm{r}+\eta_{\bm{r}}\bm{e}_\mu$. 
We also define a U(1) gauge field $A_{\bm{r},\bm{r}+\eta_{\bm{r}}\bm{e}_\mu}$ such that $S^{\pm}_i=e^{\pm i\eta_{\bm{r}}\bm{A}_{\bm{r},\bm{r}+\eta_{\bm{r}}\bm{e}_\mu}}$ and $[B_{\bm{r}_1,\bm{r}_1+\eta_{\bm{r}_1}\bm{e}_{\mu_1}},A_{\bm{r}_2,\bm{r}_2+\eta_{\bm{r}_2}\bm{e}_{\mu_2}}]=i\delta_{12}$.
They satisfy $B_{\bm{r},\bm{r}+\eta_{\bm{r}}\bm{e}_\mu}=-B_{\bm{r}+\eta_{\bm{r}}\bm{e}_\mu,\bm{r}}$ and $A_{\bm{r},\bm{r}+\eta_{\bm{r}}\bm{e}_\mu}=-A_{\bm{r}+\eta_{\bm{r}}\bm{e}_\mu,\bm{r}}$. 
Quantities that satisfy such anti-symmetric relation can be regarded as the components $\eta_{\bm{r}}\bm{e}_\mu\cdot\bm{v}(\bm{r})$ of an underlying vector field $\bm{v}(\bm{r})$. 
Now, we can introduce the U(1) LGT by substituting these definitions to $H_{\text{eff}}$,
\begin{equation}
    H_{\text{U(1)}}=\frac{U}{2}\sum_{\bm{r},\mu}\left(E_{\bm{r},\bm{r}+\eta_{\bm{r}}\bm{e}_\mu}-\frac{\eta_{\bm{r}}}{2}\right)^2-K\sum_{\hex}\cos\left(\operatorname{curl}A\right),
\end{equation}
where $\operatorname{curl}A=\sum_{\braket{\bm{r}\bm{r}'}\in\hex}A_{\bm{r}\bm{r}'}$ is the discrete curl of the U(1) gauge field. The sum over links $\braket{\bm{r}\bm{r}'}$ on the hexagon is along the direction such that these links are connected head-by-tail.

To expose the electric monopole degrees of freedom, we utilize the standard duality transformation and map the U(1) LGT obtained above to a dual U(1) LGT defined on a dual diamond lattice. The dual diamond lattice is defined by connecting the nearest-neighbor volume units (a structure in the diamond lattice consisting of 10 sites and 3 adjacent hexagons). The links of the original diamond lattice are one-to-one mapped to the volume units of the dual diamond, and vice versa. Thus, we define a dual U(1) gauge field $a_{\bm{R},\bm{R}+\eta_{\bm{r}}\bm{e}_\mu}$ on the dual diamond lattice link, such that its curl generates the original electric field,
\begin{equation}
    \operatorname{curl}a=\sum_{\braket{\bm{R}\bm{R}'}\in\hex^*}a_{\bm{R}\bm{R}'}=B_{\bm{r},\bm{r}+\eta_{\bm{r}}\bm{e}_\mu}-B_{\bm{r},\bm{r}+\eta_{\bm{r}}\bm{e}_\mu}^0.
\end{equation}
where $\hex^*$ refers to the elementary hexagon on the dual diamond lattice and the magnetic field vector $B_{\bm{r},\bm{r}+\eta_{\bm{r}}\bm{e}_\mu}$ penetrates through the center of $\hex^*$. The orientation of $\hex^*$ is defined as the direction of $\eta_\mu\bm{e}_\mu$, according to the right-hand rule. The dual diamond lattice sites are denoted by capital symbols $\bm{R}$ and $\bm{R}'$. $B_{\bm{r},\bm{r}+\eta_{\bm{R}}\bm{e}_\mu}^0$ is a background magnetic field that takes care of the ``two-in-two-out'' structure. In our convention, 
\begin{equation}
    \begin{aligned}
        & B_{\bm{r},\bm{r}+\eta_{\bm{r}}\bm{e}_0}^0=B_{\bm{r},\bm{r}+\eta_{\bm{r}}\bm{e}_1}^0=\eta_{\bm{r}}, \\
        & B_{\bm{r},\bm{r}+\eta_{\bm{r}}\bm{e}_2}^0=B_{\bm{r},\bm{r}+\eta_{\bm{r}}\bm{e}_3}^0=0.
    \end{aligned}
\end{equation}
We also define an electric field $E_{\bm{R},\bm{R}+\eta_\mu\bm{e}_\mu}$ on the dual diamond lattice links,
\begin{equation}
    E_{\bm{R},\bm{R}+\eta_\mu\bm{e}_\mu}=\operatorname{curl}A.
\end{equation}
The electric field is conjugate to the dual U(1) gauge field, $[E_{\bm{R}_1,\bm{R}_1+\eta_{\bm{R}_1}\bm{e}_{\mu_1}},a_{\bm{R}_2,\bm{R}_2+\eta_{\bm{R}_2}\bm{e}_{\mu_2}}]=i\delta_{12}$. Then, we obtain the dual U(1) LGT
\begin{equation}
    H_{\text{dual}}=\frac{U}{2}\sum_{\hex^*}\left(\operatorname{curl}a-\bar{B}\right)^2-K\sum_{\bm{R},\mu}\cos E_{\bm{R}\bm{R}'}.
\end{equation}
The sum over $\hex^*$ assumes that $\hex^*$ has a positive orientation. We also define $\bar{B}_{\bm{r},\bm{r}+\eta_{\bm{r}}\bm{e}_\mu}=B_{\bm{r},\bm{r}+\eta_{\bm{r}}\bm{e}_\mu}^0-\eta_{\bm{r}}/2$. Then, we introduce a monopole field $\theta_{\bm{R}}$ on the dual diamond lattice site that is minimally coupled with the dual U(1) gauge field~\cite{PhysRevB.94.205107}. Finally, we arrive at the following theory
\begin{equation}
    \begin{aligned}
        H_{\text{dual}}&=\frac{U}{2}\sum_{\hex^*}\left(\operatorname{curl}a-\bar{B}\right)^2-K\sum_{\bm{R},\mu}\cos E_{\bm{R}\bm{R}'}\\
        &-t\sum_{\bm{R},\mu}\cos\left(\theta_{\bm{R}}-\theta_{\bm{R}+\eta_{\bm{R}}\bm{e}_\mu}+2\pi a_{\bm{R},\bm{R}+\eta_{\bm{R}}\bm{e}_\mu}\right),
    \end{aligned}
\end{equation}
where coupling constant $t>0$ and rotor operator $\Phi_{\bm{R}}^\dagger=e^{i\theta_{\bm{R}}}$ ($\Phi_{\bm{R}}=e^{-i\theta_{\bm{R}}}$) annihilates (creates) an electric monopole on the dual diamond lattice site.

In the confinement phase, the $a$ field is slowly fluctuating. Thus, we can ignore the fluctuating of $a$ and approximate $a$ by its saddle-point value $\bar{a}$, which satisfies $\operatorname{curl}\bar{a}=\bar{B}$. This leads to a mean-field tight-binding model of the electric monopoles,
\begin{equation}
    H_{\text{m}}=-t\sum_{\braket{\bm{R}\bm{R}'}}e^{-2\pi i \bar{a}_{\bm{R}\bm{R}'}}\Phi^\dagger_{\bm{R}}\Phi_{\bm{R}'}-\mu\sum_{\bm{R}}\Phi^\dagger_{\bm{R}}\Phi_{\bm{R}}.
\end{equation}
The chemical potential $\mu$ is added to avoid the monopole condensation, and $\mu$ is of the same energy scale as the monopole bandwidth. As both the monopole and the gauge photon are operating within the spin-ice manifold, the energy scale for monopoles has to be set by the interactions that preserve the system within the spin-ice manifold. For example, these interactions include the ring exchange and further neighbor Ising interactions. The monopole Hamiltonian in the main text is derived after renaming $\overline{a}_{\bm{R}\bm{R}'}$ as $\alpha_{\bm{r}\bm{r}'}$ and $\bm{R}$ as $\bm{r}$.

The above derivation also provides the relation between the INS experiments and the three excitations. The application of $S^x_i$ or $S^y_i$ creates two spinons in the neighboring tetrahedral centers. Thus, $\braket{S^x_i S^x_j}$ or $\braket{S^y_i S^y_j}$ are directly related to the spinon bilinear, giving rise to the form factor after the Fourier transform. As for the electric monopoles, they couple with neutron spin by a loop current operator~\cite{PhysRevB.96.195127}
\begin{equation}
    J_{\bm{R}\bm{R}'}=i\left[\Phi^\dagger_{\bm{R}}\Phi_{\bm{R}'}e^{-i2\pi\bar{a}_{\bm{R}\bm{R}'}}-\text{h.c.}\right].
\end{equation}
The loop current operator is related with $S^z$ under the duality transformation, $S^z\sim B= \sum_{\braket{\bm{R}\bm{R}'}\in\hex^*}J_{\bm{R}\bm{R}'}$. Thus, $\braket{S^z_i S^z_j}$ contains the contribution from electric monopoles. After a Fourier transformation, $\braket{S^z_i S^z_j}$ yields the two-monopole density of states and the form factor. The gauge photon arises from the emergent (``magnetic'') field correlator where the emergent field here is directly coupled to the neutron spins linearly. This understanding allows one to obtain the corresponding form factors and the contributions of three excitations to INS experiments.

\bibliography{refs}